\documentclass[letterpaper]{article} 

\usepackage{aaai2026}  
\usepackage{times}  
\usepackage{helvet}  
\usepackage{courier}  
\usepackage[hyphens]{url}  
\usepackage{graphicx} 
\usepackage{natbib}  
\usepackage{caption} 
\usepackage{booktabs} 
\usepackage{multirow}
\usepackage{colortbl,xcolor}
\usepackage{tablefootnote}
\usepackage{threeparttable}
\usepackage{algorithm}
\usepackage{algorithmic}

\usepackage{tablefootnote}
\usepackage{threeparttable}

\usepackage{algorithm}
\usepackage{algorithmic}

\usepackage{multirow}
\usepackage{multicol}
\usepackage{amsmath}

\usepackage{booktabs}
\usepackage{amssymb}
\usepackage{makecell}
\usepackage{pdfpages}

\usepackage{newfloat}
\usepackage{listings}
\DeclareCaptionStyle{ruled}{labelfont=normalfont,labelsep=colon,strut=off} 
\floatstyle{ruled}
\newfloat{listing}{tb}{lst}{}
\floatname{listing}{Listing}
\title{A Scalable Pipeline for Enabling Non-Verbal Speech Generation and Understanding}
\author {
    Runchuan Ye\textsuperscript{\rm 1}\thanks{Work conducted when the first author was an intern at ModelBest.},
    Yixuan Zhou\textsuperscript{\rm 1},
    Renjie Yu\textsuperscript{\rm 1},
    Zijian Lin\textsuperscript{\rm 1},\\
    Kehan Li\textsuperscript{\rm 1},
    Xiang Li\textsuperscript{\rm 1},
    Xin Liu\textsuperscript{\rm 2},
    Guoyang Zeng\textsuperscript{\rm 2},
    Zhiyong Wu\textsuperscript{\rm 1}
}
\affiliations {
    \textsuperscript{\rm 1} Shenzhen International Graduate School, Tsinghua University, Shenzhen, China\\
    \textsuperscript{\rm 2} ModelBest Inc, Beijing, China 
}

\usepackage{bibentry}

\begin{document}

\maketitle

\begin{abstract}
Non-verbal Vocalizations (NVs), such as laughter and sighs, are vital for conveying emotion and intention in human speech, yet most existing speech systems neglect them, which severely compromises communicative richness and emotional intelligence. Existing methods for NVs acquisition are either costly and unscalable (relying on manual annotation/recording) or unnatural (relying on rule-based synthesis). To address these limitations, we propose a highly scalable automatic annotation framework to label non-verbal phenomena from natural speech, which is low-cost, easily extendable, and inherently diverse and natural. This framework leverages a unified detection model to accurately identify NVs in natural speech and integrates them with transcripts via temporal-semantic alignment method. Using this framework, we created and released \textbf{NonVerbalSpeech-38K}, a diverse, real-world dataset featuring 38,718 samples across 10 NV categories collected from in-the-wild media. Experimental results demonstrate that our dataset provides superior controllability for NVs generation and achieves comparable performance for NVs understanding. 
\end{abstract}


\begin{links}
    \link{Demos}{https://nonverbalspeech38k.github.io/nonverspeech38k/}
    \link{Datasets}{https://huggingface.co/datasets/nonverbalspeech/nonverbalspeech38k}
\end{links}

\section{Introduction}
\label{sec:intro}

Spoken interaction has become a core interface in human-computer communication, driven by the rapid advancement of automatic speech recognition (ASR), text-to-speech (TTS) synthesis, and large language models. These technologies have enabled systems to generate spoken content that is semantically coherent and impressively fluent, while also accurately understanding semantic input
~\cite{zhang2023speechgpt,xie2024mini,Moshi}.
However, human communication is far richer than just words. 
Non-verbal Vocalizations (NVs), such as laughter and sighs, are critical components of human spoken communication, carrying essential emotional, intentional, and social cues beyond lexical content
\cite{truong2007automatic,cortes2021effects,cowen2019mapping,lima2014ear}. 
Despite their importance, most current speech systems lack the ability to perceive or express such NVs, hindering the creation of truly human-like interactions.

\begin{figure}[t!]
\centering
\includegraphics[width=0.92\columnwidth]{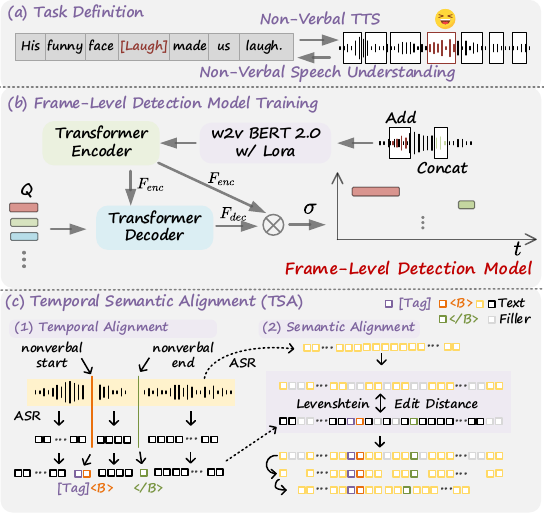}
\caption{(a) The task definition, (b) The proposed frame-level detection model, and (c) The TSA method used in the pipeline (Fig.~\ref{fig:pipeline}).
}
\label{fig:task_frame_tsa}
\end{figure}

To genuinely emulate human communication, a speech system should both recognize and produce NVs. This work focuses on two tasks capturing these capabilities: NV speech generation and understanding (Fig.~\ref{fig:task_frame_tsa} (a)).

Speech synthesis with NVs can be categorized into two distinct approaches. The first is tag-controlled, which inserts explicit tags (e.g., \texttt{[laughing]}, \texttt{[breath]}) into text to control the type and position of NVs~\cite{du2024cosyvoice2,dia}. These methods offer fine-grained control but require annotated training data and still struggle with precise timing and occurrence control. The second is spontaneous-style, which predicts NVs directly from contextual cues~\cite{li2024spontaneous}. While generally producing more natural results, such models also rely heavily on proprietary annotated data, limiting further research. 

\begin{table*}[htbp]
\centering
\footnotesize
\renewcommand{\arraystretch}{1}
\setlength{\tabcolsep}{1.1pt}
\caption{
Comparison of Existing Corpora and NVS.
}
\label{tab:dataset_comparison}
\begin{threeparttable}
\begin{tabular}{l c c c c c c}
\toprule
\textbf{Dataset} 
& \textbf{Lang.} 
& \textbf{\# Utter.} 
& \textbf{Source} 
& \textbf{Method} 
& \textbf{\# Types} 
& \textbf{Human?} \\
\midrule

NonVerbalTTS\cite{borisov2025nonverbaltts}
& EN 
& $\sim$6K 
& VoxCeleb, Expresso 
& Manual annotation
& 10
& Y \\

CapSpeech\cite{wang2025capspeech} 
& EN 
& $\sim$37K 
& LibriTTS-R 
& \makecell{Rule-based \\concatenation / overlay} 
& 3
& N \\

SMIIP-NV\cite{wu2025smiip} 
& ZH 
& $\sim$17K 
& $-$ 
& Manual recording
& 3
& Y \\

MNV-17\cite{mai2025mnv} 
& ZH 
& $\sim$3K 
& $-$ 
& Manual recording
& 17
& Y \\

SynParaSpeech\cite{bai2025synparaspeech}
& ZH 
& $\sim$56K 
& $-$ 
& \makecell{Rule-based synthesis \\with manual verification}
& 6
& Y \\

NVSpeech\cite{liao2025nvspeech}
& ZH 
& $\sim$174K 
& \makecell{Nonspeech7k,Emilia,\\Genshin,StarRail}
& \makecell{Manual labeling followed \\by model annotation}
& 18
& Y \\

\midrule
\rowcolor{gray!20}
\textbf{NVS (Ours)} 
& {EN, ZH} 
& $\sim$38K 
& {In-the-wild} 
& Fully automatic annotation 
& 10 
& {N} \\
\bottomrule
\end{tabular}
\begin{tablenotes}
\footnotesize
\item {Unlike existing works, the NVS dataset is built using a fully automated annotation pipeline, requiring no human effort and operating directly on in-the-wild data, which allows for straightforward scaling. ``\textbf{\# Types}" denotes the number of distinct non-verbal types. ``\textbf{Human?}" indicates whether human effort is required. }
\end{tablenotes}
\end{threeparttable}
\end{table*}

Beyond speech synthesis, understanding NVs is equally important for building truly human-like speech interaction systems. Traditionally, speech processing systems have focused on single-domain tasks such as ASR or sound event classification independently. However, in real-world scenarios, speech and NV sounds naturally co-occur, driving a shift toward unified models that can handle both simultaneously. Recent speech understanding models 
~\cite{tang2024salmonn, Qwen-Audio, chu2024qwen2, wu2025stepaudio2technicalreport, ding2025kimi}
employ multi-task training strategies to jointly perform ASR, sound event classification, and related speech tasks. Although this unified approach integrates multiple tasks within a single framework, effectively combining these capabilities—particularly transcribing NVs information embedded within speech transcription—remains challenging~\cite{10317236,10389855}. Progress is further constrained by the scarcity of large, well-annotated datasets.

As outlined above, the development of both NV speech synthesis and understanding remains fundamentally hindered by the scarcity of publicly available datasets. To address this, several efforts have attempted to incorporate NVs into speech datasets.
However, existing datasets often face significant limitations.
Specifically, some works rely on manual annotation or human recording, which severely restricts speaker diversity and the scale of the dataset~\cite{mai2025mnv,wu2025smiip,borisov2025nonverbaltts}. 
Other approaches use rule-based data augmentation, which may lead to issues with speech naturalness and timbre inconsistency~\cite{wang2025capspeech,bai2025synparaspeech}. 
Recent approaches that first annotate a seed dataset and then train a model for automatic expansion \cite{liao2025nvspeech} still rely heavily on costly human effort for the initial labeling.

\begin{figure*}[htbp]
\centering
\includegraphics[width=0.95\textwidth]{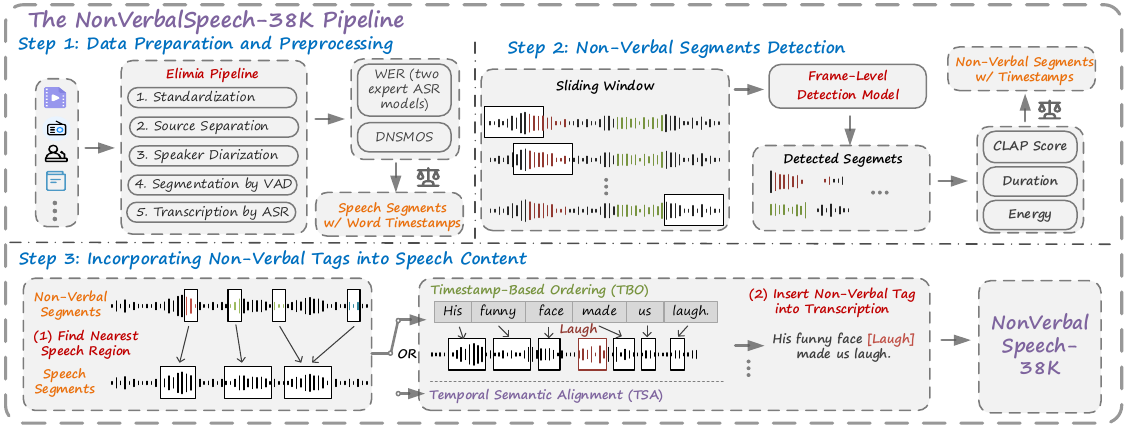}
\caption{The proposed \textit{NonVerbalSpeech-38K} data processing pipeline. Colors other than black in the waveform indicate non-verbal vocalizations.}
\label{fig:pipeline}
\end{figure*}

To address these limitations, we propose a \textbf{fully automatic annotation framework} (Fig.~\ref{fig:pipeline}) that annotates NVs from in-the-wild audio without introducing costly human effort. 
Building on this framework, we introduce \textbf{NonVerbalSpeech-38K (NVS)}, a large-scale dataset containing naturally occurring NVs along with NV-aware transcriptions.
Collected from diverse in-the-wild sources such as animations, movies, variety shows, and radio dramas, NVS contain 38K samples ($\sim$131 hours) covering 10 NV types, including \texttt{[coughing]} and \texttt{[laughing]}, as detailed in Fig.~\ref{fig:lan_and_label_dis}.
Table~\ref{tab:dataset_comparison} highlights the key differences between our dataset and existing resources. 
The key distinctions lie in three aspects: (1) our dataset supports both English and Chinese, covering a broader linguistic scope; (2) its construction is fully automatic, eliminating costly human annotation; and (3) it is derived directly from in-the-wild audio, allowing the dataset to scale continuously.

We evaluate our dataset by fine-tuning F5-TTS~\cite{chen2024f5} for NV synthesis and Qwen2-Audio~\cite{chu2024qwen2} and Whisper-Large-V3~\cite{radford2023robust} for end-to-end transcription of both verbal content and NVs. Results show that our dataset offers better controllability for NV generation and achieves competitive performance in NV speech transcription. Demos are available online\footnote{\url{https://nonverbalspeech38k.github.io/nonverspeech38k/}}.

Our main contributions are as follows:
\begin{itemize}
    \item A practical, scalable pipeline with two novel components—a unified NV detection model and a precise TSA module—for automatically and accurately detecting and labeling NV expressions from in-the-wild audio.
    \item A large-scale, diverse, well-annotated, real-world dataset covering 10 NV categories for expressive NV TTS and precise NV understanding has been released\footnote{\url{https://huggingface.co/datasets/nonverbalspeech/nonverbalspeech38k}}.
    \item Fine-tuning F5-TTS with our NVS (TSA) consistently outperforms the strongest baseline (CLAP score $0.110$ (EN) and $0.100$ (ZH); IMOS $3.22$ (EN) and $0.33$ (ZH)) (Table~\ref{tab:tts_main_result}), achieving relative gains of $70\%$, $79\%$, $8\%$, and $11\%$, respectively, showing superior NV controllability.
    \item Fine-tuning Qwen2-Audio and Whisper-Large-V3 with our NVS (TSA) reduces TPD (sec.~\ref{subsubsec:nv_caption_eval}) by 0.92 and 2.64, respectively, compared with the baseline (Table~\ref{tab:exp_asr_tag_en}), indicating enhanced NV perception.
\end{itemize}

\section{Methodology}

This section presents our \textit{NonVerbalSpeech-38K} annotation pipeline (Fig.~\ref{fig:pipeline}). It consists of three main stages: Data Preparation and Preprocessing, NV Segment Detection, and Incorporation of NV Tags into Speech Content. 
We describe each stage in detail below.
\subsection{Data Preparation and Preprocessing}

As shown in Fig.~\ref{fig:pipeline} (Step 1), we first gather in-the-wild data from various sources, such as movies and cartoons. Using the Emilia-Pipeline~\cite{he2024emilia}, we preprocess the data through standardization, source separation, and speaker diarization to extract single-speaker segments. Voice activity detection (VAD) removes silence, followed by transcription with Whisper-Large-V3~\cite{radford2023robust} and Paraformer-zh~\cite{gao2022paraformer}. Segments with WER~$>20\%$ or DNSMOS~$<1.0$ are filtered to ensure high-quality speech. This process also yields segment- and word-level timestamps for further processing.

\subsection{Non-Verbal Segments Detection}
In this step, we first train the unified frame-level NV detection model and then apply it within the pipeline.

\subsubsection{Model Architecture}

Inspired by T-UAED \cite{jiang2025unified}, we propose a unified frame-level detection model to localize NV events, as illustrated in Fig.~\ref{fig:task_frame_tsa} (b). 
Specifically, we employ a pretrained wav2vec-bert 2.0 model 
~\cite{barrault2023seamless}
to extract $L$-layer features $F_i \in \mathbb{R}^{B \times T \times D}$. To effectively utilize multi-level representations, we apply a learnable weighted sum:
\begin{equation}
F = \sum_{i=1}^{L} \alpha_i F_i, \quad \text{s.t. } \sum \alpha_i = 1,\ \alpha_i \geq 0
\end{equation}

The fused representation $F$ is then passed through a Transformer encoder, followed by a Transformer decoder with learnable queries $Q \in \mathbb{R}^{B \times N_q \times D}$:
\begin{equation}
F_{\text{enc}} = \text{Encoder}(F), \quad F_{\text{dec}} = \text{Decoder}(Q, F_{\text{enc}})
\end{equation}

To obtain frame-level predictions, we compute the similarity between decoder and encoder outputs:
\begin{equation}
S = \sigma(F_{\text{dec}} F_{\text{enc}}^\top), \quad S \in \mathbb{R}^{B \times N_q \times T}
\end{equation}

Here, $\sigma(\cdot)$ denotes the sigmoid function. Finally, the model is trained using binary cross-entropy loss.

\subsubsection{Data Preparation for Model Training}
To train the frame-level detection model, we construct augmented samples by concatenating or overlaying NVs and speech segments (Fig. \ref{fig:task_frame_tsa} (b)).
NV clips are collected from widely used sound event datasets including VocalSound
~\cite{gong_vocalsound}
, VGGSound
~\cite{chen2020vggsound}
, Nonspeech7k
~\cite{rashid2023nonspeech7k}
, AudioSet-Strong
~\cite{hershey2021benefit}
, and ESC-50
~\cite{piczak2015esc}
.
To ensure quality, we first apply energy-based VAD to remove silence and split each NV clip into multiple segments. Segments with low CLAP scores ($<$ 0.4) or very short durations ($<$ 0.3s) are discarded. 
The resulting high-quality segments are used to augment training samples, with their distribution shown in Fig.~\ref{fig:frame_model_test_result_and_nv_samples}.
Speech data is drawn from the GigaSpeech M subset ($\sim$1,000 hrs) 
\cite{chen2021gigaspeech}
and SEED-TTS-Eval 
\cite{anastassiou2024seed}
. The GigaSpeech test split and the SEED-TTS-Eval set are used exclusively for evaluation.

\subsubsection{Experimental Setup and Results}

\begin{figure}[tbhp]
\centering
\includegraphics[width=1.0\columnwidth]{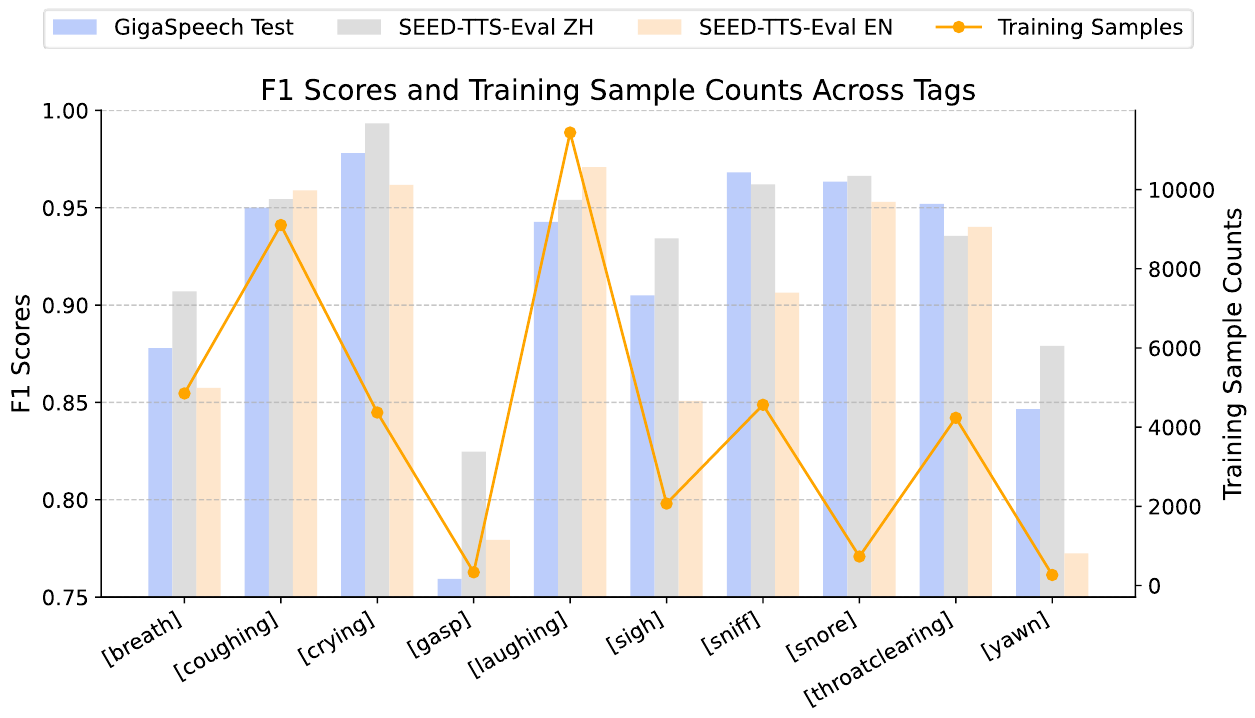}
\caption{Frame-level detection model test results and distribution of training NV segments.}
\label{fig:frame_model_test_result_and_nv_samples}
\end{figure}

We set $N_q = 10$, corresponding to the 10 NV types in Fig.~\ref{fig:frame_model_test_result_and_nv_samples}. Both encoder and decoder have 6 layers. 
The model is fine-tuned with LoRA~\cite{hu2022lora} for 300 epochs on 8 H100 GPUs, using a batch size of 64 per GPU. The learning rate is set to $1 \times 10^{-4}$, with a 5\% warm-up followed by cosine annealing.
See Appendix for details.
After training, the model achieves a frame-level F1-score of around 91\% on three test splits (Fig.~\ref{fig:frame_model_test_result_and_nv_samples}).
With a similar number of training samples, \texttt{[gasp]} is relatively difficult to perceive and detect, whereas \texttt{[snore]} and \texttt{[yawn]} are easier to distinguish due to their distinctive acoustic patterns. Similarly, \texttt{[breath]} performs slightly worse than \texttt{[crying]} and \texttt{[throatclearing]}.
Overall, the frame-level detection model performs relatively poorly on \texttt{[gasp]} and \texttt{[yawn]} due to limited training samples, but performs well on other NV types. 
Notably, even when trained solely on English speech, the model performs well on Chinese test data, demonstrating strong language-agnostic capabilities and potential for scalability.

\begin{table*}[!t]
\centering
\caption{
Evaluation results of NV speech generation on English (EN) and Chinese (ZH) test sets.
}
\footnotesize
\renewcommand{\arraystretch}{1}
\setlength{\tabcolsep}{1.2pt}
\label{tab:tts_main_result}
\begin{threeparttable}
\begin{tabular}{l|cc|cc|cc|cc|cc|cc}
\toprule
\multirow{2}{*}{\textbf{Model}}
& \multicolumn{2}{c|}{\textbf{CLAP Score} $\uparrow$}
& \multicolumn{2}{c|}{\textbf{WER/CER (\%)} $\downarrow$}
& \multicolumn{2}{c|}{\textbf{DNSMOS} $\uparrow$}
& \multicolumn{2}{c|}{\textbf{SSIM} $\uparrow$}
& \multicolumn{2}{c|}{\textbf{QMOS} $\uparrow$}
& \multicolumn{2}{c}{\textbf{IMOS} $\uparrow$} \\
& EN & ZH & EN & ZH & EN & ZH & EN & ZH & EN & ZH & EN & ZH \\
\midrule
\textcolor{gray}{CosyVoice2} & \textcolor{gray}{$0.087$} & \textcolor{gray}{$0.071$} & \textcolor{gray}{$3.941$} & \textcolor{gray}{$5.012$} & \textcolor{gray}{${3.236}$} & \textcolor{gray}{${3.280}$} & \textcolor{gray}{$0.514$} & \textcolor{gray}{$0.671$} & \textcolor{gray}{$3.58_{(\pm 0.06)}$} & \textcolor{gray}{$3.54_{(\pm 0.07)}$} & \textcolor{gray}{$2.12_{(\pm 0.06)}$} & \textcolor{gray}{$2.31_{(\pm 0.08)}$} \\
\textcolor{gray}{Dia} & \textcolor{gray}{$0.182$} & \textcolor{gray}{$0.097$} & \textcolor{gray}{$15.687$} & \textcolor{gray}{$171.882$} & \textcolor{gray}{$2.749$} & \textcolor{gray}{$2.528$} & \textcolor{gray}{$0.431$} & \textcolor{gray}{$0.319$} & \textcolor{gray}{$3.47_{(\pm 0.13)}$} & \textcolor{gray}{$-$} & \textcolor{gray}{$3.27_{(\pm 0.14)}$} & \textcolor{gray}{$-$} \\
\midrule
\textcolor{gray}{F5-TTS} & \textcolor{gray}{$-$} & \textcolor{gray}{$-$} &
\textcolor{gray}{{$1.700$}} & \textcolor{gray}{{$2.755$}} & \textcolor{gray}{$3.086$} & \textcolor{gray}{$3.238$} & \textcolor{gray}{{$0.659$}} & \textcolor{gray}{$0.713$} & \textcolor{gray}{$3.72_{(\pm 0.09)}$} & \textcolor{gray}{$3.27_{(\pm 0.10)}$} & \textcolor{gray}{$-$} & \textcolor{gray}{$-$} \\
\hline
\quad + CapSpeech & \textcolor{gray}{$0.125$} & \textcolor{gray}{$0.121$} 
& \underline{$2.791$} & $9.856$ & $3.009$ & $3.153$ & $0.570$ & $0.669$ & $3.66_{(\pm 0.10)}$ & $2.40_{(\pm 0.07)}$ & \textcolor{gray}{$1.79_{(\pm 0.07)}$} & \textcolor{gray}{$1.88_{(\pm 0.08)}$} \\
\quad + NVTTS & $0.110$ & $0.095$ & $\mathbf{2.279}$ & $7.042$ & $3.099$ & $\underline{3.248}$ & $0.566$ & $0.631$ & $\mathbf{3.80}_{(\pm 0.08)}$ & $3.28_{(\pm 0.08)}$ & $3.22_{(\pm 0.10)}$ & $3.33_{(\pm 0.08)}$ \\
\quad + NVSpeech & \textcolor{gray}{$0.092$} & \textcolor{gray}{$0.094$} 
& $3.525$ & $4.726$ & $3.040$ & $3.214$ & $\mathbf{0.596}$ & $\underline{0.683}$ & $3.49_{(\pm 0.10)}$ & $3.32_{(\pm 0.07)}$ & \textcolor{gray}{$2.83_{(\pm 0.11)}$} & \textcolor{gray}{$3.36_{(\pm 0.09)}$}\\
\quad + SMIIP-NV & \textcolor{gray}{$0.257$} & \textcolor{gray}{$0.281$} 
& ${3.501}$ & ${3.910}$ & $\mathbf{3.132}$ & $\mathbf{3.259}$ & $0.497$ & $0.604$ & $3.75_{(\pm 0.08)}$ & $3.39_{(\pm 0.08)}$ & \textcolor{gray}{$3.47_{(\pm 0.13)}$} & \textcolor{gray}{$3.57_{(\pm 0.11)}$} \\
\quad + SynParaSpeech & \textcolor{gray}{$0.077$} & \textcolor{gray}{$0.057$} 
& $3.605$ & $4.605$ & $3.084$ & $3.192$ & {$0.589$} & $\mathbf{0.685}$ & $3.25_{(\pm 0.08)}$ & $3.00_{(\pm 0.07)}$ & \textcolor{gray}{$2.88_{(\pm 0.12)}$} & \textcolor{gray}{$3.12_{(\pm 0.09)}$} \\
\quad + MNV-17 & $0.087$ & $0.100$ & $5.457$ & $4.670$ & $3.102$ & $\underline{3.248}$ & $\underline{0.590}$ & $0.667$ & $3.43_{(\pm 0.11)}$ & $\underline{3.41}_{(\pm 0.08)}$ & $2.79_{(\pm 0.10)}$ & $3.22_{(\pm 0.07)}$ \\
\hline
\rowcolor{gray!20}
\quad + NVS(TBO) & \underline{$0.113$} & \underline{$0.110$} & $3.561$ & \underline{$3.216$} & $3.079$ & $3.239$ & $0.565$ & $0.668$ & $3.72_{(\pm 0.08)}$ & ${3.38}_{(\pm 0.08)}$ & $\underline{3.26}_{(\pm 0.07)}$ & $\underline{3.48}_{(\pm 0.07)}$ \\
\rowcolor{gray!20}
\quad + NVS(TSA) & $\mathbf{0.187}$ & $\mathbf{0.179}$ & $3.498$ & $\mathbf{3.101}$ & \underline{$3.114$} & $3.236$ & $0.567$ & $0.670$ & $\underline{3.76}_{(\pm 0.08)}$ & $\mathbf{3.43}_{(\pm 0.09)}$ & $\mathbf{3.48}_{(\pm 0.08)}
$ & $\mathbf{3.71}_{(\pm 0.07)}$ \\
\bottomrule
\end{tabular}
\begin{tablenotes}
    \item The best and second-best values in each column are highlighted in \textbf{bold} and \underline{underline}, respectively. Note that CosyVoice2, Dia, and F5-TTS are excluded as reference models. Gray entries in CLAP Score and QMOS indicate that only a subset of NV types is supported and are therefore excluded from the ranking, see Table~\ref{tab:clap_break_down} for a more detailed comparison. NVTTS: NonVerbalTTS.
\end{tablenotes}
\end{threeparttable}
\end{table*}

\subsubsection{Non-Verbal Region Detection Using the Trained Frame-Level Detection Model}

After training the frame-level detection model, we apply it to standardized waveforms using a sliding window strategy, as shown in Fig. \ref{fig:pipeline} (Step 2). 
Standardized waveforms are used because tools like Silero-VAD often miss NV events at the beginning or end of speech segments.
To provide sufficient context for detection, we use a window size of 10s and a hop size of 5s.
To improve detection quality, we apply three filters: (1) discard segments shorter than 0.3\,s; (2) remove segments with very low energy ($<-35\,\mathrm{dB}$), such as subtle breathing sounds; and (3) discard segments with a CLAP score below 0.3, since higher scores indicate true NV events. These thresholds are determined based on a simple inspection of the data.

\subsection{Incorporating Non-Verbal Tags into Speech Content}

After filtering out low-quality candidates, we match each NV segment with the nearest speech region obtained from preprocessing (Fig.~\ref{fig:pipeline}(Step 3(1))). The pseudocode for this process is provided in the Appendix.
It should be noted that NV events generally do not lie entirely within speech regions. Therefore, any NV segment located more than 1 second away from the nearest speech region is discarded. 

After this, we obtain high-quality speech segments containing NV events, together with NV labels and their start/end timestamps, as well as speech transcriptions with word-level timestamps.  
To support NV speech generation and understanding, it is necessary to integrate the textual content with NV tags. 
As shown in Step 3 of Fig.~\ref{fig:pipeline}, for a segment transcribed as “His funny face made us laugh,” a laughter sound occurs right after “His funny face.” By aligning their timestamps, we obtain the final transcript: “His funny face [laugh] made us laugh.” The algorithm is provided in the Appendix.
However, the timestamp-based ordering (TBO) method assumes accurate word timestamps. In practice, ASR timestamps often contain offsets, causing misalignment between the TBO-generated NV-aware transcription and the speech.
The Montreal Forced Aligner was expected to improve time stamp precision, but the resulting alignments show greater temporal drift.

To address this, we propose the Temporal-Semantic Alignment (TSA) method (Fig.~\ref{fig:task_frame_tsa}, (c)), yielding two dataset variants (TBO and TSA). We first cut the speech at NV boundaries, run ASR on each segment, and merge the results with NV tags to obtain timing-accurate NV-aware transcriptions. Since this step may introduce semantic inconsistencies, we further align the NV-aware output with the full-segment ASR transcription using edit-distance alignment, combining accurate NV timings with coherent semantics. The final annotation integrates NV tags with the full-segment ASR output. NV events rarely overlap with semantic content, with \texttt{<B>} and \texttt{</B>} co-occurring in less than 2\% of cases in the NVS dataset. 
Note that TSA does not rely on any word-level time stamps, which is a key advantage over TBO.

\section{The NonVerbalSpeech-38K Dataset}

\begin{figure}[tbhp]
\centering
\includegraphics[width=0.85\columnwidth]{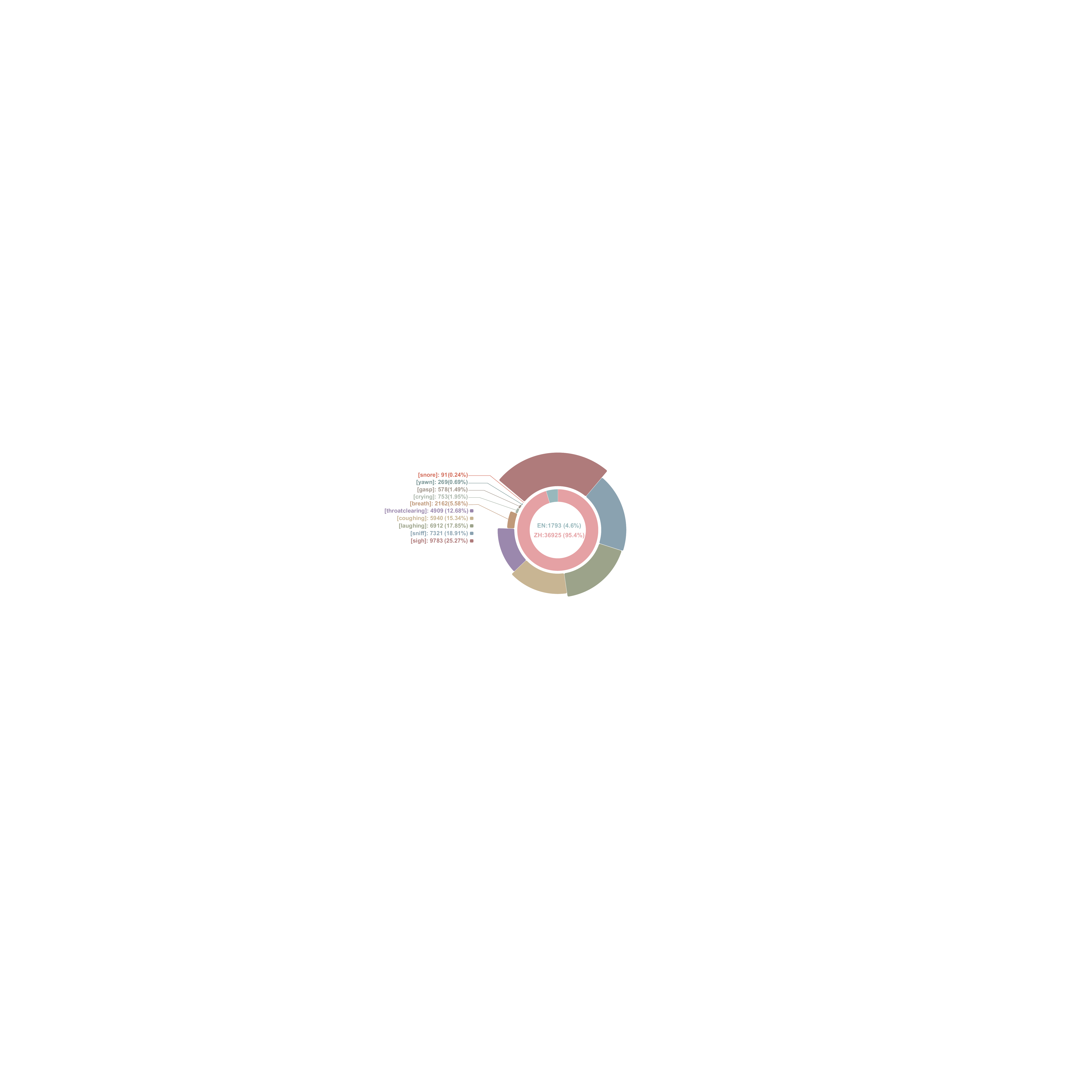}
\caption{Language and NV label distribution in proposed NVS dataset.}
\label{fig:lan_and_label_dis}
\end{figure}

Using the proposed NonVerbalSpeech-38K pipeline, we construct a dataset from diverse online speech sources (statistics are provided in the Appendix), covering a wide range of natural speaking styles and rich NVs. 
The initial release contains $\sim$131 hours of English and Chinese speech, totaling about 38,000 samples. As shown in Fig.~\ref{fig:lan_and_label_dis}, Chinese samples dominate due to the composition of the source data. Again, the frame-level detection model trained only on English still generalizes well to Chinese, indicating strong cross-lingual scalability.
Frequent NV types such as \texttt{[sigh]}, \texttt{[sniff]}, \texttt{[laughing]}, and \texttt{[coughing]} dominate, reflecting their prevalence in daily speech. Less frequent types (e.g., \texttt{[crying]}, \texttt{[yawn]}) are also included. 
Further information and analysis can be found in the Appendix.

\begin{table}[t!]
\centering
\scriptsize
\begin{threeparttable}
\renewcommand{\arraystretch}{1}
\setlength{\tabcolsep}{1.2pt}
\setlength{\extrarowheight}{0pt}
\caption{CLAP Score Differences (Comparison Systems $-$ F5-TTS Fine-Tuned with NVS (TSA)), Broken Down by Non-Verbal Tag.}
\begin{tabular}{lcccccc}
\toprule
\textbf{Model} & \textbf{[laughing]} & \textbf{[throat]} & \textbf{[sigh]} & \textbf{[coughing]} & \textbf{[sniff]} & \textbf{[breath]} \\
\midrule
\multicolumn{7}{c}{\textit{English (EN)}} \\
\midrule
CosyVoice2 & $-0.132$ & $-$ & $-0.149$ & $-0.150$ & $-$ & $-0.056$ \\
Dia & $-0.050$ & $-0.034$ & $\mathbf{0.004}$ & $\mathbf{0.062}$ & $-0.085$ & $-$ \\
F5-TTS & $-$ & $-$ & $-$ & $-$ & $-$ & $-$ \\
\quad + CapSpeech & $-0.132$ & $-$ & $-$ & $-0.151$ & $-$ & $-0.015$ \\
\quad + NVTTS & $-0.160$ & $-0.003$ & $-0.136$ & $-0.043$ & $-0.043$ & $-0.077$ \\
\quad + NVSpeech & $-0.131$ & $-$ & $-0.135$ & $-0.144$ & $-$ & $-0.058$ \\
\quad + SMIIP-NV & $-0.072$ & $-$ & $-$ & $\mathbf{0.027}$ & $-$ & $-$ \\
\quad + SynParaSpeech & $-0.117$ & $-0.116$ & $-0.071$ & $-$ & $-$ & $-$ \\
\quad + MNV-17 & $-0.171$ & $-0.084$ & $-0.128$ & $-0.142$ & $-0.047$ & $-0.032$ \\
\hline
\rowcolor{gray!20}
\quad + NVS(TBO) & $-0.076$ & $-0.043$ & $-0.105$ & $-0.099$ & $-0.074$ & $-0.051$ \\
\midrule
\multicolumn{7}{c}{\textit{Chinese (ZH)}} \\
\midrule
CosyVoice2 & $-0.134$ & $-$ & $-0.124$ & $-0.152$ & $-$ & $-0.098$ \\
Dia & $-0.105$ & $\mathbf{0.004}$ & $-0.056$ & $-0.132$ & $-0.150$ & $-$ \\
F5-TTS & $-$ & $-$ & $-$ & $-$ & $-$ & $-$ \\
\quad + CapSpeech & $-0.157$ & $-$ & $-$ & $-0.127$ & $-$ & $-0.039$ \\
\quad + NVTTS & $-0.146$ & $-0.044$ & $-0.108$ & $-0.058$ & $-0.070$ & $-0.078$ \\
\quad + NVSpeech & $-0.140$ & $-$ & $-0.103$ & $-0.118$ & $-$ & $-0.055$ \\
\quad + SMIIP-NV & $-0.029$ & $-$ & $-$ & $\mathbf{0.058}$ & $-$ & $-$ \\
\quad + SynParaSpeech & $-0.120$ & $-0.113$ & $-0.060$ & $-$ & $-$ & $-$ \\
\quad + MNV-17 & $-0.180$ & $-0.052$ & $-0.092$ & $-0.105$ & $-0.015$ & $-0.033$ \\
\hline
\rowcolor{gray!20}
\quad + NVS(TBO) & $-0.060$ & $-0.042$ & $-0.101$ & $-0.109$ & $-0.057$ & $-0.051$ \\
\bottomrule
\end{tabular}
\label{tab:clap_break_down}
\begin{tablenotes}
\footnotesize
\item Negative values indicate that F5-TTS fine-tuned with NVS (TSA) outperforms the corresponding system.
    \texttt{[throat]} denotes \texttt{[throatclearing]}, and ``$-$'' indicates unsupported NV types for the current comparison system.
\end{tablenotes}
\end{threeparttable}
\end{table}

\section{Experiments}
We evaluate the effectiveness of our dataset through two downstream tasks: NV speech generation and NV speech understanding. The details are as follows.

\subsection{Experimental Setup for Non-Verbal Speech Generation}

\subsubsection{Implementation Details}
In the NV speech generation task, we use F5-TTS as the base model. 
To support NV control, we expand the vocabulary of F5-TTS by adding NV tags such as \texttt{[breath]}, \texttt{[sigh]}.
We fine-tune F5-TTS on seven datasets (Table~\ref{tab:dataset_comparison}), including the two variants of our NVS dataset. See Appendix for details.

\subsubsection{Baselines}

In addition to fine-tuning F5-TTS on various datasets, we include two strong baselines, CosyVoice2~\cite{du2024cosyvoice2} and Dia~\cite{dia}, both supporting NV control and generation.
 
\subsubsection{Evaluation}
Due to differences in NV support across models and datasets, we select six common NV phenomena for evaluation (Table~\ref{tab:clap_break_down}).
To evaluate NV speech generation, we use the prompt speech from the seed-tts-eval framework~\cite{anastassiou2024seed}.
For text inputs, tagged sentences are generated using Gemini~\cite{team2023gemini} and GPT-4o~\cite{GPT-4o}.
For each NV tag, we construct 175 sentences in both Chinese and English, yielding 1,050 sentences per language.
To evaluate system performance, we use standard metrics including DNSMOS, SSIM, and WER/CER to assess speech quality, speaker similarity, and intelligibility. Additionally, we introduce the CLAP Score to measure the presence of intended NV phenomena in synthesized speech.
For subjective evaluation, 20 participants rated the samples on QMOS (for overall speech quality) and IMOS (for accuracy in rendering both verbal content and NV tags). See Appendix for details.

\subsection{Experimental Setup for Non-Verbal Speech Understaning}

\subsubsection{Implementation Details}
We use Qwen2-Audio and Whisper-Large-V3 as base models, neither of which supports NV-aware transcription. To enable this, we add special NV tokens and fine-tune the models on CapSpeech, NVSpeech, and both NVS variants. See Appendix for details.

\subsubsection{Evaluation}
\label{subsubsec:nv_caption_eval}

To evaluate the model's ability to integrate NV tags into transcriptions, 
we use a subset of NonVerbalTTS as the English test set (754 samples, containing \texttt{[laughing]}, \texttt{[breath]}, and \texttt{[coughing]}), and a subset of SMIIP-NV and MNV-17 as the Chinese test set (784 samples, containing \texttt{[laughing]}, \texttt{[sigh]}, and \texttt{[coughing]}). 
All three datasets are manually constructed, ensuring high annotation quality (Table~\ref{tab:dataset_comparison}).
In addition to standard WER/CER (excluding NV tags) and NV-aware WER/CER (NVWER/NVCER, including them), we use two tag-level metrics further evaluating position accuracy: \textbf{Tag Position Distance (TPD)} and its normalized version \textbf{Normalized Tag Distance (NTD)}. TPD and NTD are computed after aligning predicted and reference transcriptions using Levenshtein edit distance, where
\begin{equation}
\text{TPD} = \frac{1}{n} \sum_{i=1}^{n} |p_i - g_i|, \quad
\text{NTD} = \frac{\text{TPD}}{L},
\end{equation}
with $p_i$ and $g_i$ denoting matched tag positions in the prediction and reference and $L$ the reference length. Lower values indicate better accuracy. See Appendix for details.

\section{Experimental Results}

\subsection{Experimental Results on Non-Verbal TTS}

Table~\ref{tab:tts_main_result} summarizes NV speech generation results for English and Chinese.
Among baselines, CosyVoice2 excels in speech quality (highest DNSMOS and ZH QMOS) but shows limited NV controllability (low CLAP-Score and IMOS). Dia performs well on EN NV metrics (CLAP Score and IMOS) but suffers from poor stability and high failure rates, and is unsupported in Chinese, leading to the exclusion of its ZH subjective scores.
F5-TTS is a strong base model, achieving the highest SSIM and lowest WER/CER. However, the pretrained F5-TTS does not support NV generation.

Fine-tuning on NVS (TSA) achieves superior Chinese performance, with top subjective scores, strong objective metrics, and the highest CLAP Score, indicating competitive intelligibility, naturalness, and NV controllability.
CapSpeech and NVTTS underperform in ZH CER due to English-only training data.
On English tests, NVS(TSA) also performs well, achieving the highest IMOS and CLAP Score, and second-highest DNSMOS and QMOS.
CapSpeech attains a high CLAP Score but lowest DNSMOS and IMOS, likely due to unnatural synthetic non-verbal sounds.
Tables~\ref{tab:clap_break_down} present CLAP Score comparisons per NV tag. We observe that NVS (TSA) ranks below existing models in only a few cases, demonstrating comprehensive and stable NV controllability. A similar trend is also observed for IMOS, see the Appendix for more details.

Overall, NVS (TSA) performs well across QMOS, IMOS, CLAP Score, ZH CER, and DNSMOS. Compared with NVS (TBO), it shows significant improvements in CLAP Score and IMOS, mainly because TSA provides more accurate annotations, particularly regarding NV positions, highlighting the importance of precise NV labels for controllable NV generation.
Nonetheless, the NVS dataset exhibits some limitations in SSIM and English WER, likely due to its size, which we plan to address by further expanding the dataset in future work.

\begin{table}[t!]
\centering
\footnotesize
\renewcommand{\arraystretch}{1}
\setlength{\tabcolsep}{0.5pt}
\caption{
EN Evaluation results for NV speech understanding.}
\begin{tabular}{lcccc}
\toprule
\textbf{Model} 
& \textbf{WER (\%)} $\downarrow$ 
& \textbf{NVWER (\%)} $\downarrow$ 
& \textbf{TPD} $\downarrow$  
& \textbf{NTD} $\downarrow$ \\
\midrule
\textcolor{gray}{Qwen2-Audio}
& \textcolor{gray}{33.087} 
& \textcolor{gray}{--} 
& \textcolor{gray}{--} 
& \textcolor{gray}{--} \\
\hline
\quad + CapSpeech & 20.065 & 23.261 & 4.232 & 0.178 \\
\rowcolor{gray!20}
\quad + NVS(TBO) & \textbf{16.901} & \textbf{21.952} & 6.223 & 0.291 \\
\rowcolor{gray!20}
\quad + NVS(TSA) & 17.035 & 22.099 & \textbf{3.314} & \textbf{0.159} \\
\hline
\textcolor{gray}{Whisper-Large-V3}
& \textcolor{gray}{17.542} 
& \textcolor{gray}{--} 
& \textcolor{gray}{--} 
& \textcolor{gray}{--} \\
\hline
\quad + CapSpeech  & 16.037 & 19.780 & 3.380 & 0.167 \\

\rowcolor{gray!20}
\quad + NVS(TBO) & \textbf{15.469} & {21.843} & 8.490 & 0.396 \\

\rowcolor{gray!20}
\quad + NVS(TSA) & 15.503 & \textbf{19.661} & \textbf{0.740} & \textbf{0.055} \\
\bottomrule
\end{tabular}
\label{tab:exp_asr_tag_en}
\end{table}

\begin{table}[t]
\centering
\footnotesize
\renewcommand{\arraystretch}{1}
\setlength{\tabcolsep}{1pt}
\caption{
ZH Evaluation results for NV speech understanding.}
\begin{tabular}{lcccc}
\toprule
\footnotesize
\textbf{Model} 
& \textbf{CER (\%)} $\downarrow$ 
& \textbf{NVCER (\%)} $\downarrow$ 
& \textbf{TPD} $\downarrow$  
& \textbf{NTD} $\downarrow$ \\
\midrule

\textcolor{gray}{Qwen2-Audio}
& \textcolor{gray}{13.868} 
& \textcolor{gray}{--} 
& \textcolor{gray}{--} 
& \textcolor{gray}{--} \\
\hline
\quad + NVSpeech
& 4.314 & 4.994 & 1.530 & 0.050 \\

\rowcolor{gray!20}
\quad + NVS(TBO) 
& 6.002 & 8.467 & 5.419 & 0.182 \\

\rowcolor{gray!20}
\quad + NVS(TSA)
& 4.921 & 6.093 & 1.737 & 0.062 \\
\hline
\textcolor{gray}{Whisper-Large-V3}
& \textcolor{gray}{6.225} 
& \textcolor{gray}{--} 
& \textcolor{gray}{--} 
& \textcolor{gray}{--} \\
\hline
\quad + NVSpeech 
& 4.841 & 5.166 & 0.904 & 0.030 \\

\rowcolor{gray!20}
\quad + NVS(TBO) 
& 5.864 & 8.141 & 4.526 & 0.170 \\

\rowcolor{gray!20}
\quad + NVS(TSA)
& 5.087 & 5.844 & 1.110 & 0.038 \\

\bottomrule
\end{tabular}
\label{tab:exp_asr_tag_zh}
\end{table}

\subsection{Experimental Results on Non-Verbal Speech Understaning}

Table~\ref{tab:exp_asr_tag_en} and \ref{tab:exp_asr_tag_zh} compares NV speech understanding performance.
On the ZH test set, NVS (TSA) demonstrates competitive performance despite the smaller dataset size and fully automatic construction method without human effort (Table~\ref{tab:dataset_comparison}), with all metrics remaining low.
On the EN test set, it consistently outperforms CapSpeech, which is constructed via rule-based synthesis and limited by dataset diversity.
Notably, TSA shows significant improvements over TBO in TPD and NTD across both ZH and EN settings, further validating the superiority of the TSA method, consistent with Table~\ref{tab:tts_main_result}.

\section{Discussion}
Finally, we highlight the scalability of our pipeline:
\begin{itemize}
    \item The unified NV detection model is language-agnostic, can generalize to other sounds by increasing query tokens $N_q$, and, with CLAP text embeddings, can also support description-based, fine-grained sound detection.
    \item Unlike prior work that relies on human effort, our pipeline is fully automated, suggesting that the dataset can be further scaled—a direction for future work. Unlike rule-based synthesis methods, it captures natural NV sounds from diverse sources, which improves NV modeling.
\end{itemize}

\section{Conclusion}

In this work, we design a fully automated NV annotation framework featuring two novel components—a unified detection model and the TSA alignment module—avoiding reliance on costly manual labeling and the unnatural outputs caused by rule-based synthesis in prior work.
Using this framework, we directly annotate NV phenomena from in-the-wild speech, creating and publicly releasing the NVS dataset, which contains diverse and natural NV events with precise textual labels. 
Experimental results demonstrate that this dataset effectively enabling existing models to perform NV generation and recognition, particularly exhibiting superior NV controllability.

\bibliography{aaai2026}

@article{xie2024mini,
  title={Mini-omni: Language models can hear, talk while thinking in streaming},
  author={Xie, Zhifei and Wu, Changqiao},
  journal={arXiv preprint arXiv:2408.16725},
  year={2024}
}

@inproceedings{zhang2023speechgpt,
  title={SpeechGPT: Empowering Large Language Models with Intrinsic Cross-Modal Conversational Abilities},
  author={Zhang, Dong and Li, Shimin and Zhang, Xin and Zhan, Jun and Wang, Pengyu and Zhou, Yaqian and Qiu, Xipeng},
  booktitle={Findings of the Association for Computational Linguistics: EMNLP 2023},
  pages={15757--15773},
  year={2023}
}

@inproceedings{hershey2021benefit,
  title={The benefit of temporally-strong labels in audio event classification},
  author={Hershey, Shawn and Ellis, Daniel PW and Fonseca, Eduardo and Jansen, Aren and Liu, Caroline and Moore, R Channing and Plakal, Manoj},
  booktitle={ICASSP 2021-2021 IEEE International Conference on Acoustics, Speech and Signal Processing (ICASSP)},
  pages={366--370},
  year={2021},
  organization={IEEE}
}

@INPROCEEDINGS{10317236,
  author={Shione, Nagito and Wakabayashi, Yukoh and Kitaoka, Norihide},
  booktitle={2023 Asia Pacific Signal and Information Processing Association Annual Summit and Conference (APSIPA ASC)}, 
  title={Construction of Automatic Speech Recognition Model that Recognizes Linguistic Information and Verbal/Non-verbal Phenomena}, 
  year={2023},
  volume={},
  number={},
  pages={2306-2311},
  doi={10.1109/APSIPAASC58517.2023.10317236}}

@INPROCEEDINGS{10389855,
  author={Shione, Nagito and Wakabayashi, Yukoh and Kitaoka, Norihide},
  booktitle={2023 10th International Conference on Advanced Informatics: Concept, Theory and Application (ICAICTA)}, 
  title={Automatic Speech Recognition Using Linguistic and Verbal/Non-Verbal Information}, 
  year={2023},
  volume={},
  number={},
  pages={1-6},
  doi={10.1109/ICAICTA59291.2023.10389855}}

@article{wang2025capspeech,
  title={CapSpeech: Enabling Downstream Applications in Style-Captioned Text-to-Speech},
  author={Wang, Helin and Hai, Jiarui and Chong, Dading and Thakkar, Karan and Feng, Tiantian and Yang, Dongchao and Lee, Junhyeok and Velazquez, Laureano Moro and Villalba, Jesus and Qin, Zengyi and others},
  journal={arXiv preprint arXiv:2506.02863},
  year={2025}
}

@inproceedings{piczak2015esc,
  title={ESC: Dataset for environmental sound classification},
  author={Piczak, Karol J},
  booktitle={Proceedings of the 23rd ACM international conference on Multimedia},
  pages={1015--1018},
  year={2015}
}

@inproceedings{chen2020vggsound,
  title={Vggsound: A large-scale audio-visual dataset},
  author={Chen, Honglie and Xie, Weidi and Vedaldi, Andrea and Zisserman, Andrew},
  booktitle={ICASSP 2020-2020 IEEE International Conference on Acoustics, Speech and Signal Processing (ICASSP)},
  pages={721--725},
  year={2020},
  organization={IEEE}
}

@article{borisov2025nonverbaltts,
  title={NonverbalTTS: A Public English Corpus of Text-Aligned Nonverbal Vocalizations with Emotion Annotations for Text-to-Speech},
  author={Borisov, Maksim and Spirin, Egor and Diatlova, Daria},
  journal={arXiv preprint arXiv:2507.13155},
  year={2025}
}

@article{du2024cosyvoice2,
  title={Cosyvoice 2: Scalable streaming speech synthesis with large language models},
  author={Du, Zhihao and Wang, Yuxuan and Chen, Qian and Shi, Xian and Lv, Xiang and Zhao, Tianyu and Gao, Zhifu and Yang, Yexin and Gao, Changfeng and Wang, Hui and others},
  journal={arXiv preprint arXiv:2412.10117},
  year={2024}
}

@inproceedings{li2024spontaneous,
  title={Spontaneous Style Text-to-Speech Synthesis with Controllable Spontaneous Behaviors Based on Language Models},
  author={Li, Weiqin and Yang, Peiji and Zhong, Yicheng and Zhou, Yixuan and Wang, Zhisheng and Wu, Zhiyong and Wu, Xixin and Meng, Helen},
  booktitle={Proc. Interspeech 2024},
  pages={1785--1789},
  year={2024}
}

@article{Qwen-Audio,
  title={Qwen-Audio: Advancing Universal Audio Understanding via Unified Large-Scale Audio-Language Models},
  author={Chu, Yunfei and Xu, Jin and Zhou, Xiaohuan and Yang, Qian and Zhang, Shiliang and Yan, Zhijie  and Zhou, Chang and Zhou, Jingren},
  journal={arXiv preprint arXiv:2311.07919},
  year={2023}
}

@article{chu2024qwen2,
  title={Qwen2-audio technical report},
  author={Chu, Yunfei and Xu, Jin and Yang, Qian and Wei, Haojie and Wei, Xipin and Guo, Zhifang and Leng, Yichong and Lv, Yuanjun and He, Jinzheng and Lin, Junyang and others},
  journal={arXiv preprint arXiv:2407.10759},
  year={2024}
}

@article{ding2025kimi,
  title={Kimi-audio technical report},
  author={Ding, Ding and Ju, Zeqian and Leng, Yichong and Liu, Songxiang and Liu, Tong and Shang, Zeyu and Shen, Kai and Song, Wei and Tan, Xu and Tang, Heyi and others},
  journal={arXiv preprint arXiv:2504.18425},
  year={2025}
}

@inproceedings{tang2024salmonn,
  title={{SALMONN}: Towards Generic Hearing Abilities for Large Language Models},
  author={Changli Tang and Wenyi Yu and Guangzhi Sun and Xianzhao Chen and Tian Tan and Wei Li and Lu Lu and Zejun MA and Chao Zhang},
  booktitle={The Twelfth International Conference on Learning Representations},
  year={2024},
  url={https://openreview.net/forum?id=14rn7HpKVk}
}

@misc{wu2025stepaudio2technicalreport,
      title={Step-Audio 2 Technical Report},
      author={Boyong Wu and Chao Yan and Chen Hu and Cheng Yi and Chengli Feng and Fei Tian and Feiyu Shen and Gang Yu and Haoyang Zhang and Jingbei Li and Mingrui Chen and Peng Liu and Wang You and Xiangyu Tony Zhang and Xingyuan Li and Xuerui Yang and Yayue Deng and Yechang Huang and Yuxin Li and Yuxin Zhang and Zhao You and Brian Li and Changyi Wan and Hanpeng Hu and Jiangjie Zhen and Siyu Chen and Song Yuan and Xuelin Zhang and Yimin Jiang and Yu Zhou and Yuxiang Yang and Bingxin Li and Buyun Ma and Changhe Song and Dongqing Pang and Guoqiang Hu and Haiyang Sun and Kang An and Na Wang and Shuli Gao and Wei Ji and Wen Li and Wen Sun and Xuan Wen and Yong Ren and Yuankai Ma and Yufan Lu and Bin Wang and Bo Li and Changxin Miao and Che Liu and Chen Xu and Dapeng Shi and Dingyuan Hu and Donghang Wu and Enle Liu and Guanzhe Huang and Gulin Yan and Han Zhang and Hao Nie and Haonan Jia and Hongyu Zhou and Jianjian Sun and Jiaoren Wu and Jie Wu and Jie Yang and Jin Yang and Junzhe Lin and Kaixiang Li and Lei Yang and Liying Shi and Li Zhou and Longlong Gu and Ming Li and Mingliang Li and Mingxiao Li and Nan Wu and Qi Han and Qinyuan Tan and Shaoliang Pang and Shengjie Fan and Siqi Liu and Tiancheng Cao and Wanying Lu and Wenqing He and Wuxun Xie and Xu Zhao and Xueqi Li and Yanbo Yu and Yang Yang and Yi Liu and Yifan Lu and Yilei Wang and Yuanhao Ding and Yuanwei Liang and Yuanwei Lu and Yuchu Luo and Yuhe Yin and Yumeng Zhan and Yuxiang Zhang and Zidong Yang and Zixin Zhang and Binxing Jiao and Daxin Jiang and Heung-Yeung Shum and Jiansheng Chen and Jing Li and Xiangyu Zhang and Yibo Zhu},
      year={2025},
      eprint={2507.16632},
      archivePrefix={arXiv},
      primaryClass={cs.CL},
      url={https://arxiv.org/abs/2507.16632},
}

@inproceedings{jiang2025unified,
  title={Unified audio event detection},
  author={Jiang, Yidi and Tao, Ruijie and Huang, Wen and Chen, Qian and Wang, Wen},
  booktitle={ICASSP 2025-2025 IEEE International Conference on Acoustics, Speech and Signal Processing (ICASSP)},
  pages={1--5},
  year={2025},
  organization={IEEE}
}

@misc{dia,
  author       = {Nari-labs},
  title        = {Dia},
  year         = {2025},
  howpublished = {https://github.com/nari-labs/dia},
}

@article{GPT-4o,
  title={Gpt-4o system card},
  author={Hurst, Aaron and Lerer, Adam and Goucher, Adam P and Perelman, Adam and Ramesh, Aditya and Clark, Aidan and Ostrow, AJ and Welihinda, Akila and Hayes, Alan and Radford, Alec and others},
  journal={arXiv preprint arXiv:2410.21276},
  year={2024}
}

@article{Moshi,
  title={Moshi: a speech-text foundation model for real-time dialogue},
  author={D{\'e}fossez, Alexandre and Mazar{\'e}, Laurent and Orsini, Manu and Royer, Am{\'e}lie and P{\'e}rez, Patrick and J{\'e}gou, Herv{\'e} and Grave, Edouard and Zeghidour, Neil},
  journal={arXiv preprint arXiv:2410.00037},
  year={2024}
}

@article{chen2024f5,
  title={F5-tts: A fairytaler that fakes fluent and faithful speech with flow matching},
  author={Chen, Yushen and Niu, Zhikang and Ma, Ziyang and Deng, Keqi and Wang, Chunhui and Zhao, Jian and Yu, Kai and Chen, Xie},
  journal={arXiv preprint arXiv:2410.06885},
  year={2024}
}

@article{anastassiou2024seed,
  title={Seed-tts: A family of high-quality versatile speech generation models},
  author={Anastassiou, Philip and Chen, Jiawei and Chen, Jitong and Chen, Yuanzhe and Chen, Zhuo and Chen, Ziyi and Cong, Jian and Deng, Lelai and Ding, Chuang and Gao, Lu and others},
  journal={arXiv preprint arXiv:2406.02430},
  year={2024}
}

@inproceedings{gao2022paraformer,
  title={Paraformer: Fast and Accurate Parallel Transformer for Non-autoregressive End-to-End Speech Recognition},
  author={Gao, Zhifu and Zhang, ShiLiang and McLoughlin, Ian and Yan, Zhijie},
  booktitle={Proc. Interspeech 2022},
  pages={2063--2067},
  year={2022}
}

@INPROCEEDINGS{gong_vocalsound,
  author={Gong, Yuan and Yu, Jin and Glass, James},
  booktitle={ICASSP 2022 - 2022 IEEE International Conference on Acoustics, Speech and Signal Processing (ICASSP)}, 
  title={Vocalsound: A Dataset for Improving Human Vocal Sounds Recognition}, 
  year={2022},
  pages={151-155},
  doi={10.1109/ICASSP43922.2022.9746828}}

@article{rashid2023nonspeech7k,
  title={Nonspeech7k dataset: Classification and analysis of human non-speech sound},
  author={Rashid, Muhammad Mamunur and Li, Guiqing and Du, Chengrui},
  journal={IET Signal Processing},
  volume={17},
  number={6},
  pages={e12233},
  year={2023},
  publisher={Wiley Online Library}
}

@inproceedings{hu2022lora,
title={Lo{RA}: Low-Rank Adaptation of Large Language Models},
author={Edward J Hu and yelong shen and Phillip Wallis and Zeyuan Allen-Zhu and Yuanzhi Li and Shean Wang and Lu Wang and Weizhu Chen},
booktitle={International Conference on Learning Representations},
year={2022},
url={https://openreview.net/forum?id=nZeVKeeFYf9}
}

@inproceedings{chen2021gigaspeech,
  title={GigaSpeech: An Evolving, Multi-Domain ASR Corpus with 10,000 Hours of Transcribed Audio},
  author={Chen, Guoguo and Chai, Shuzhou and Wang, Guan-Bo and Du, Jiayu and Zhang, Wei-Qiang and Weng, Chao and Su, Dan and Povey, Daniel and Trmal, Jan and Zhang, Junbo and others},
  booktitle={Proc. Interspeech 2021},
  pages={3670--3674},
  year={2021}
}

@inproceedings{he2024emilia,
  title={Emilia: An extensive, multilingual, and diverse speech dataset for large-scale speech generation},
  author={He, Haorui and Shang, Zengqiang and Wang, Chaoren and Li, Xuyuan and Gu, Yicheng and Hua, Hua and Liu, Liwei and Yang, Chen and Li, Jiaqi and Shi, Peiyang and others},
  booktitle={2024 IEEE Spoken Language Technology Workshop (SLT)},
  pages={885--890},
  year={2024},
  organization={IEEE}
}

@inproceedings{wu2023large,
  title={Large-scale contrastive language-audio pretraining with feature fusion and keyword-to-caption augmentation},
  author={Wu, Yusong and Chen, Ke and Zhang, Tianyu and Hui, Yuchen and Berg-Kirkpatrick, Taylor and Dubnov, Shlomo},
  booktitle={ICASSP 2023-2023 IEEE International Conference on Acoustics, Speech and Signal Processing (ICASSP)},
  pages={1--5},
  year={2023},
  organization={IEEE}
}

@article{truong2007automatic,
  title={Automatic discrimination between laughter and speech},
  author={Truong, Khiet P and Van Leeuwen, David A},
  journal={Speech Communication},
  volume={49},
  number={2},
  pages={144--158},
  year={2007},
  publisher={Elsevier}
}

@article{cowen2019mapping,
  title={Mapping 24 emotions conveyed by brief human vocalization.},
  author={Cowen, Alan S and Elfenbein, Hillary Anger and Laukka, Petri and Keltner, Dacher},
  journal={American psychologist},
  volume={74},
  number={6},
  pages={698},
  year={2019},
  publisher={American Psychological Association}
}

@article{lima2014ear,
  title={In the ear of the beholder: how age shapes emotion processing in nonverbal vocalizations.},
  author={Lima, C{\'e}sar F and Alves, Tiago and Scott, Sophie K and Castro, S{\~a}o Lu{\'\i}s},
  journal={Emotion},
  volume={14},
  number={1},
  pages={145},
  year={2014},
  publisher={American Psychological Association}
}

@article{cortes2021effects,
  title={Effects of aging on emotion recognition from dynamic multimodal expressions and vocalizations},
  author={Cortes, Diana S and Tornberg, Christina and B{\"a}nziger, Tanja and Elfenbein, Hillary Anger and Fischer, H{\aa}kan and Laukka, Petri},
  journal={Scientific reports},
  volume={11},
  number={1},
  pages={2647},
  year={2021},
  publisher={Nature Publishing Group UK London}
}

@article{team2023gemini,
  title={Gemini: a family of highly capable multimodal models},
  author={Team, Gemini and Anil, Rohan and Borgeaud, Sebastian and Alayrac, Jean-Baptiste and Yu, Jiahui and Soricut, Radu and Schalkwyk, Johan and Dai, Andrew M and Hauth, Anja and Millican, Katie and others},
  journal={arXiv preprint arXiv:2312.11805},
  year={2023}
}

@inproceedings{radford2023robust,
  title={Robust speech recognition via large-scale weak supervision},
  author={Radford, Alec and Kim, Jong Wook and Xu, Tao and Brockman, Greg and McLeavey, Christine and Sutskever, Ilya},
  booktitle={International conference on machine learning},
  pages={28492--28518},
  year={2023},
  organization={PMLR}
}

@inproceedings{mcauliffe2017montreal,
  title={Montreal forced aligner: Trainable text-speech alignment using kaldi.},
  author={McAuliffe, Michael and Socolof, Michaela and Mihuc, Sarah and Wagner, Michael and Sonderegger, Morgan},
  booktitle={Interspeech},
  volume={2017},
  pages={498--502},
  year={2017}
}

@article{lee2022bigvgan,
  title={Bigvgan: A universal neural vocoder with large-scale training},
  author={Lee, Sang-gil and Ping, Wei and Ginsburg, Boris and Catanzaro, Bryan and Yoon, Sungroh},
  journal={arXiv preprint arXiv:2206.04658},
  year={2022}
}

@inproceedings{wu2025smiip,
  title={SMIIP-NV: A Multi-Annotation Non-Verbal Expressive Speech Corpus in Mandarin for LLM-Based Speech Synthesis},
  author={Wu, Zhuojun and Liu, Dong and Liu, Juan and Wang, Yechen and Li, Linxi and Jin, Liwei and Bu, Hui and Zhang, Pengyuan and Li, Ming},
  booktitle={Proceedings of the 33rd ACM International Conference on Multimedia},
  pages={12564--12570},
  year={2025}
}

@article{mai2025mnv,
  title={MNV-17: A High-Quality Performative Mandarin Dataset for Nonverbal Vocalization Recognition in Speech},
  author={Mai, Jialong and Ji, Jinxin and Xing, Xiaofen and Yang, Chen and Chen, Weidong and Xing, Jingyuan and Xu, Xiangmin},
  journal={arXiv preprint arXiv:2509.18196},
  year={2025}
}

@article{bai2025synparaspeech,
  title={SynParaSpeech: Automated Synthesis of Paralinguistic Datasets for Speech Generation and Understanding},
  author={Bai, Bingsong and Lu, Qihang and Yang, Wenbing and Sun, Zihan and Hou, Yueran and Jia, Peilei and Pu, Songbai and Fu, Ruibo and Gao, Yingming and Li, Ya and others},
  journal={arXiv preprint arXiv:2509.14946},
  year={2025}
}

@article{liao2025nvspeech,
  title={NVSpeech: An Integrated and Scalable Pipeline for Human-Like Speech Modeling with Paralinguistic Vocalizations},
  author={Liao, Huan and Ni, Qinke and Wang, Yuancheng and Lu, Yiheng and Zhan, Haoyue and Xie, Pengyuan and Zhang, Qiang and Wu, Zhizheng},
  journal={arXiv preprint arXiv:2508.04195},
  year={2025}
}

@article{barrault2023seamless,
  title={Seamless: Multilingual Expressive and Streaming Speech Translation},
  author={Barrault, Lo{\"\i}c and Chung, Yu-An and Meglioli, Mariano Coria and Dale, David and Dong, Ning and Duppenthaler, Mark and Duquenne, Paul-Ambroise and Ellis, Brian and Elsahar, Hady and Haaheim, Justin and others},
  journal={arXiv preprint arXiv:2312.05187},
  year={2023}
}


\newpage
\appendix

\makeatletter
\renewcommand{\section}{\@startsection{section}{1}{0mm}
  {\baselineskip}
  {0.5\baselineskip}
  {\normalfont\Large\bfseries\raggedright}}
\makeatother

\begin{table*}[tbph]
\centering
\small
\setlength{\tabcolsep}{4pt}
\caption{Subjective evaluation criteria for QMOS and IMOS.}
\begin{tabular}{c|p{7.3cm}|p{7.3cm}}
\toprule
\textbf{Metrics} & \textbf{QMOS: Quality Evaluation} & \textbf{IMOS: Instruction Following Evaluation} \\
\midrule
Guidance & 
Please rate each audio sample based on \textit{naturalness} (whether it sounds like real human speech), \textit{sound quality} (whether it is clean and undistorted), and \textit{intelligibility} (how easily the content can be understood). 
&
Please rate each audio sample based on how well it follows the given text instructions.
For example, given the text: \texttt{[sigh] Today's work is not even halfway done yet.}
The TTS system is expected not only to synthesize the text correctly but also to follow the \texttt{[sigh]} tag by producing a sigh sound at the corresponding position.
All tags used in the evaluation and their meanings are as follows:
\texttt{[laughing]}: laughing sound,
\texttt{[coughing]}: coughing sound, 
\texttt{[sigh]}: sighing sound,
\texttt{[sniff]}: sniffing sound,
\texttt{[throatclearing]}: throat clearing sound,
\texttt{[breath]}: breathing sound.
\\
\midrule
\textbf{Score} & \textbf{Criteria} & \textbf{Criteria} \\
\midrule
5 & 
Excellent naturalness, pure sound quality, and perfectly intelligible content. &
Fully follows the text and accurately renders the non-verbal tags, 
the speech sounds entirely human-like.\\
\midrule
4 & 
Good naturalness, mostly clean sound, and clear content. &
Fully follows the text, and non-verbal tags are rendered naturally. \\
\midrule
3 & 
Moderate naturalness, minor sound artifacts (e.g., slight noise), and mostly intelligible content. &
Mostly follows the text; non-verbal tag rendering is generally accurate but lacks naturalness. \\
\midrule
2 & 
Poor naturalness, noticeable noise or distortion, and hard-to-understand content. &
Slight omissions of text, or missing tags, or incorrectly rendered tags (e.g., tags appearing in incorrect positions or associated with the wrong sounds). \\
\midrule
1 & 
Very unnatural (e.g., mechanical), heavily distorted, and nearly unintelligible. &
Significant text loss or synthesis failure (e.g., unintelligible or missing content). \\
\bottomrule
\end{tabular}

\label{tab:qmos_imos_criteria}
\end{table*}

\begin{table*}[htbp]
\centering
\begin{threeparttable}
\footnotesize
\caption{Counts of non-verbal tags in the training and test sets for the English setting.}
\setlength{\tabcolsep}{12pt}
\renewcommand{\arraystretch}{1.3}
\begin{tabular}{lcccc}
\toprule
\textbf{Dataset} & \textbf{[laughing]} & \textbf{[coughing]} & \textbf{[breath]} & \textbf{Total} \\
\midrule
CapSpeech (Train) & 19,571 & 15,174 & 2,110 & 36,855 \\
NonVerbalSpeech-38K Subset (Train) & 6,912 & 5,940 & 2,162 & 15,014 \\
\hline
\textbf{NonVerbalTTS Subset (Test)} & \textbf{367} & \textbf{137} & \textbf{250} & \textbf{754} \\
\bottomrule
\end{tabular}
\label{tab:nonverbal_counts_for_understanding_en_train_and_test}
\begin{tablenotes}
\footnotesize
\item CapSpeech and a subset of NonVerbalSpeech-38K are used for training, while the NonVerbalTTS subset is used for testing.
\end{tablenotes}
\end{threeparttable}
\end{table*}

\begin{table*}[htbp]
\centering
\begin{threeparttable}
    
\footnotesize
\caption{Counts of non-verbal tags in the training and test sets for the Chinese setting.}
\setlength{\tabcolsep}{12pt}
\renewcommand{\arraystretch}{1.3}
\begin{tabular}{lcccc}
\toprule
\textbf{Dataset} & \textbf{[laughing]} & \textbf{[coughing]} & \textbf{[sigh]} & \textbf{Total} \\
\midrule
NVspeech Subset (Train) & 20,938 & 3,691 & 7,717 & 32,346 \\
NonVerbalSpeech-38K Subset (Train) & 6,912 & 5,940 & 9,783 & 22,635 \\
\hline
\textbf{SMIIP-NV\&MNV-17 Subset (Test)} & \textbf{280} & \textbf{278} & \textbf{226} & \textbf{784} \\
\bottomrule
\end{tabular}
\label{tab:nonverbal_counts_for_understanding_zh_train_and_eval}
\begin{tablenotes}
    \item  A subset of NonVerbalSpeech-38K and NVSpeech are used for training, while a subset of SMIIP-NV and a subset of MNV-17 are combined and used for testing.
\end{tablenotes}
\end{threeparttable}
\end{table*}

\section{Experimental Setup and Results}
In this section, we detail the experimental setup for the frame-level detection model, non-verbal speech generation, and non-verbal speech understanding.

\subsection{Experimental Setup and Results for Frame-level Detection Model}
We set $N_q$ to 10, corresponding to the 10 non-verbal phenomena, 
and the number of layers $L$ for the pre-trained wav2vec-bert 2.0 model \cite{barrault2023seamless} is 25. Both the encoder and decoder are configured with 6 layers, where the decoder excludes the self-attention module. The $Q$ embeddings are randomly initialized. We adopt LoRA \cite{hu2022lora} to fine-tune all linear layers within the self-attention modules of the wav2vec-bert 2.0 model. The model is trained for 300 epochs with a batch size of 64 per GPU, using 8 NVIDIA H100 GPUs. The learning rate is set to $1 \times 10^{-4}$ and scheduled with a 5\% warm-up phase followed by cosine annealing decay.

After training, the frame-level F1-Score achieves approximately 91\% across all three test splits, namely the GigaSpeech test split \cite{chen2021gigaspeech}, SEED-TTS-Eval Chinese speech \cite{anastassiou2024seed}, and SEED-TTS-Eval English speech \cite{anastassiou2024seed}. Note that these test sets are also constructed by randomly concatenating or overlaying non-verbal clips with speech segments.

\subsection{Experimental Setup and Results for Non-Verbal Speech Generation}

\subsubsection{Implementation Details}
In the non-verbal speech generation task, we use F5-TTS \cite{chen2024f5} as the base model. 
It is a fully non-autoregressive speech synthesis model that adopts a flow-matching method combined with a transformer architecture. We use the pretrained F5-TTS model, which was trained on the Chinese and English subsets of the Emilia dataset \cite{he2024emilia}, and further fine-tune it on the non-verbal dataset. 
We use BigVGAN-v2~\cite{lee2022bigvgan} as the vocoder. BigVGAN-v2 is trained on datasets that include diverse types of audio, such as speech in multiple languages, environmental sounds, and musical instruments. This diversity is important for non-verbal speech generation, as it helps the vocoder better handle a wide range of acoustic patterns.
To support non-verbal control, we expand the vocabulary of F5-TTS by adding non-verbal tags such as \texttt{[laughing]}, \texttt{[coughing]}. 
Inspired by CapSpeech~\cite{wang2025capspeech}, we initialize the embeddings of the new tokens using representations from the CLAP \cite{wu2023large} text encoder, which is trained on large-scale audio-text alignment data, including non-verbal audio-text pairs.

To retain the basic zero-shot text-to-speech ability of the pretrained model while allowing it to learn non-verbal speech generation, we carefully adjust the total number of training steps based on the scale of each dataset.
Based on empirical observations, for all datasets—including ours—but excluding NVSpeech~\cite{liao2025nvspeech}, SynParaSpeech~\cite{bai2025synparaspeech}, and CapSpeech~\cite{wang2025capspeech}, we train the models for 400 epochs to avoid overfitting and maintain fundamental TTS capability.
For CapSpeech, we train the model for 50 epochs to achieve the best performance, as extending the training to 400 epochs significantly degrades intelligibility and results in noticeably unnatural speech.
NVSpeech and SynParaSpeech are trained for approximately 60k steps, matching the number of training steps used for our dataset, which ensures a fair comparison while preventing overfitting.

All models are optimized using the AdamW optimizer with a base learning rate of $1\mathrm{e}{-5}$, a 10\% warm-up phase, and a cosine annealing learning rate schedule. For the newly added non-verbal tokens, a higher learning rate of $1\mathrm{e}{-4}$ is applied to encourage faster adaptation. Fine-tuning is performed on 8 NVIDIA H100 GPUs, with a batch size of 38,400 frames per GPU.

\subsubsection{Evaluation}
To evaluate system performance, in addition to standard metrics such as DNSMOS\footnote{\url{https://github.com/microsoft/DNS-Challenge}}, SSIM (speaker similarity)\footnote{\url{https://github.com/BytedanceSpeech/seed-tts-eval}}, and WER/CER (Word Error Rate or Character Error Rate)\footnote{\url{https://github.com/BytedanceSpeech/seed-tts-eval}} for assessing speech quality, speaker similarity, and intelligibility, respectively, we also introduce the CLAP Score~\cite{wu2023large} to evaluate whether the synthesized speech contains the intended non-verbal phenomena. 
Specifically, we use the template sentence ``\texttt{This is a sound of \textbf{specific non-verbal sound}.}'' (e.g., for laughing, ``\texttt{This is a sound of laughing.}'') as the text input to the CLAP~\cite{wu2023large} text tower, and feed the entire synthesized waveform into the CLAP audio tower.
For subjective evaluation, we adopt two metrics. The first is QMOS, which assesses the overall quality of the synthesized speech, focusing on intelligibility, audio quality, and naturalness. The second is IMOS, which evaluates whether the audio follows the input text, including non-verbal tags. For example, given the input ``\texttt{[sigh] The work for today is not even halfway done yet.}", the TTS system is expected not only to correctly synthesize the verbal content but also to emit a sigh sound at the appropriate position as indicated by the \texttt{[sigh]} tag. We collected a total of 20 subjective questionnaires. Detailed scoring criteria are provided in the table~\ref{tab:qmos_imos_criteria}.

\subsubsection{Experimental Results}

Table~\ref{tab:imos_break_down} presents IMOS comparisons across NV tags. We observe that NVS (TSA) ranks below existing models in only a few cases and achieves consistent leadership in Chinese, demonstrating comprehensive and stable NV controllability.

\subsection{Experimental Setup for Non-Verbal Speech Understaning}

\subsubsection{Implementation Details}
In this task, we use Qwen2-Audio and Whisper-Large-V3 as base models, extending their vocabularies to support non-verbal tags (e.g., [laughing], [breath]).
For Qwen2-audio, the fine-tuning is performed using LoRA, applied to the linear layers within the self-attention modules of the language model. During training, only the LoRA parameters and the newly introduced tokens are updated, while all other model parameters are frozen. 
All models are trained using a learning rate of 7.5e-5, batch size 12 per GPU, with 10\% warmup and cosine learning rate decay, using 8 NVIDIA H100 GPUs.
For Whisper-Large-V3, we full-parameters fine-tuneing the whisper decoder while keep the encoder frozen. 
All models are trained using a learning rate of 7.5e-5, batch size 24 per GPU, with 10\% warmup and cosine learning rate decay, using 8 NVIDIA H100 GPUs.

In the English setting, for a fair comparison, we use only the subset of NonVerbalSpeech-38K containing the three non-verbal tags supported by CapSpeech (i.e., \texttt{[laughing]}, \texttt{[breath]}, and \texttt{[coughing]}) for training. Details are shown in Table~\ref{tab:nonverbal_counts_for_understanding_en_train_and_test}.
Both datasets (including the two variants of our NonVerbalSpeech-38K) are trained for 5 epochs.
In the Chinese setting, we use the overlapping tags in NVSpeech and our dataset (i.e., \texttt{[laughing]}, \texttt{[coughing]}, \texttt{[sigh]}) for training. Details are shown in Table~\ref{tab:nonverbal_counts_for_understanding_zh_train_and_eval}. Both datasets (including the two variants of our NonVerbalSpeech-38K) are trained for 5 epochs. Note that both CapSpeech and NVSpeech have more total samples than our NVS dataset. Even under this setting, our dataset outperforms CapSpeech and achieves performance comparable to NVSpeech, demonstrating its efficiency and effectiveness.

\subsubsection{Evaluation}

In the English setting, we use a subset of NonVerbalTTS for evaluation, as shown in Table~\ref{tab:nonverbal_counts_for_understanding_en_train_and_test}. 
In the Chinese setting, a subset of SMIIP-NV and a subset of MNV-17 are combined and used for evaluation, as shown in Table~\ref{tab:nonverbal_counts_for_understanding_zh_train_and_eval}. 
To simplify the evaluation, we include only samples that contain a single non-verbal tag.

For the evaluation metrics, in addition to standard WER/CER (excluding non-verbal tags) and NV-aware WER/CER (NVWER/NVCER, including them), we introduce two tag-level metrics to further evaluate position accuracy: \textbf{Tag Position Distance (TPD)} and its normalized variant, \textbf{Normalized Tag Distance (NTD)}.

\begin{itemize}
    \item \textbf{Tag Position Distance (TPD):} To assess the positional accuracy of predicted non-verbal tags, we first align the predicted transcription $T^{(p)}$ with the reference transcription $T^{(r)}$ using the Levenshtein edit distance algorithm. This produces aligned sequences $T_{\text{align}}^{(p)}$ and $T_{\text{align}}^{(r)}$, 
    an example is shown in Table~\ref{tab:ref_hyp_aligned_by_edit_distance}. 
    Let $P = \{p_1, p_2, \dots, p_n\}$ and $G = \{g_1, g_2, \dots, g_n\}$ denote the positions of the $n$ aligned tags in the predicted and reference sequences, respectively. The TPD is defined as:
    \begin{equation}
    \text{TPD} = \frac{1}{n} \sum_{i=1}^{n} |p_i - g_i|.
    \end{equation}
    A lower TPD indicates better alignment of non-verbal tags. In our experiments, both the training and evaluation datasets contain only one tag per utterance to simplify the evaluation and more accurately reflect model performance.    
    \item \textbf{Normalized Tag Distance (NTD):} To reduce the influence of sequence length, TPD is normalized by the length of the aligned reference transcription:
    \begin{equation}
    \text{NTD} = \frac{\text{TPD}}{L}, \quad \text{where } L = |T_{\text{align}}^{(r)}|.
    \end{equation}
\end{itemize}

\begin{table}[tbph]
\centering
\scriptsize
\begin{threeparttable}
\renewcommand{\arraystretch}{1.1}
\setlength{\tabcolsep}{0.7pt}
\setlength{\extrarowheight}{0pt}
\caption{IMOS Differences (Comparison Systems $-$ F5-TTS Fine-Tuned with NVS (TSA)), Broken Down by Non-Verbal Tag.}
\begin{tabular}{lcccccc}
\toprule
\textbf{Model} & \textbf{[laughing]} & \textbf{[throat]} & \textbf{[sigh]} & \textbf{[coughing]} & \textbf{[sniff]} & \textbf{[breath]} \\
\midrule
\multicolumn{7}{c}{\textit{English (EN)}} \\
\midrule
CosyVoice2 & $-1.40$ & $-$ & $-1.08$ & $-1.60$ & $-$ & $-1.15$ \\
Dia & $\mathbf{0.22}$ & $\mathbf{0.53}$ & $-0.95$ & $-1.37$ & $-0.40$ & $-$ \\
F5-TTS & $-$ & $-$ & $-$ & $-$ & $-$ & $-$ \\
\quad + CapSpeech & $-1.81$ & $-$ & $-$ & $-2.03$ & $-$ & $-1.38$ \\
\quad + NVTTS & $-0.68$ & $\mathbf{0.33}$ & $-0.90$ & $\mathbf{0.03}$ & \quad + $\mathbf{0.04}$ & $-0.43$ \\
\quad + NVSpeech & $-0.38$ & $-$ & $-0.60$ & $-0.43$ & $-$ & $-0.90$ \\
\quad + SMIIP-NV & $-0.11$ & $-$ & $-$ & $-0.25$ & $-$ & $-$ \\
\quad + SynParaSpeech & $-0.46$ & $-0.62$ & $-0.30$ & $-$ & $-$ & $-$ \\
\quad + MNV-17 & $-1.35$ & $-0.10$ & $-0.10$ & $-0.13$ & $-0.96$ & $-1.22$ \\
\hline
\rowcolor{gray!20}
\quad + NVS(TBO) & $\mathbf{0.05}$ & $-0.17$ & $-0.25$ & $-0.75$ & $-0.38$ & $-0.07$ \\
\midrule
\multicolumn{7}{c}{\textit{Chinese (ZH)}} \\
\midrule
CosyVoice2 & $-0.77$ & $-$ & $-1.62$ & $-1.58$ & $-$ & $-1.78$ \\
F5-TTS & $-$ & $-$ & $-$ & $-$ & $-$ & $-$ \\
\quad + CapSpeech & $-1.50$ & $-$ & $-$ & $-2.20$ & $-$ & $-1.80$ \\
\quad + NVTTS & $-0.50$ & $-0.68$ & $-0.57$ & $-0.03$ & $-0.34$ & $-0.05$ \quad + \\
\quad + NVSpeech & $-0.47$ & $-$ & $-0.10$ & $-0.58$ & $-$ & $-0.02$ \\
\quad + SMIIP-NV & $-0.18$ & $-$ & $-$ & $-0.06$ & $-$ & $-$ \\
\quad + SynParaSpeech & $-0.13$ & $-0.60$ & $-0.67$ & $-$ & $-$ & $-$ \\
\quad + MNV-17 & $-0.83$ & $-0.26$ & $-0.60$ & $-0.53$ & $-0.32$ & $-0.25$ \\
\hline
\rowcolor{gray!20}
\quad + NVS(TBO) & $-0.03$ & $-0.20$ & $-0.62$ & $-0.05$ & $-0.40$ & $-0.25$ \\
\bottomrule
\end{tabular}
\label{tab:imos_break_down}
\begin{tablenotes}
\scriptsize
\item Negative values indicate that F5-TTS fine-tuned with NVS (TSA) outperforms the corresponding system. Due to its high CER in Chinese, Dia is excluded from the IMOS ZH evaluation.
    \texttt{[throat]} denotes \texttt{[throatclearing]}, and ``$-$'' indicates unsupported NV types for the current comparison system.
\end{tablenotes}
\end{threeparttable}
\end{table}

\begin{table*}[ht]
\centering
\caption{Example alignment between reference (REF) and model hypothesis (HYP) with non-verbal tag \texttt{[coughing]}. }
\label{tab:ref_hyp_aligned_by_edit_distance}
\begin{threeparttable}
    
\small
\renewcommand{\arraystretch}{1.2}
\begin{tabular}{@{}p{1cm}p{15cm}@{}}
\toprule
\textbf{} & \textbf{Aligned by Edit Distance} \\
\midrule
REF & \texttt{[coughing] i just did what \textbf{****} everybody else does you know after \textbf{*****} school \textbf{they} kind of \textbf{look} for the next thing and the next thing was you know a levels and} \\
HYP & \texttt{[coughing] i just did what \textbf{what} everybody else does you know after \textbf{after} school \textbf{he} kind of \textbf{looked} for the next thing and the next thing was you know a levels and} \\
\bottomrule
\end{tabular}
\begin{tablenotes}
    \item Insertions (e.g., repeated \texttt{what}, \texttt{after}) and substitutions (e.g., \texttt{he} vs. \texttt{they}) are observed.
\end{tablenotes}
\end{threeparttable}
\end{table*}

\section{NonVerbalSpeech-38K Pipeline}

\subsection{Details on Incorporating Non-Verbal Tags into Speech Content}

In this step, each non-verbal segment is matched with the nearest speech region obtained during preprocessing. The pseudocode for identifying the nearest speech region is provided in Algorithm~\ref{alg:nearest_speech}.

\begin{algorithm}[htbp]
\centering
\caption{Find Nearest Speech Region for a Non-Verbal Segment}
\label{alg:nearest_speech}
\begin{algorithmic}[1]
\REQUIRE Speech regions $S=\{s_i\}$; non-verbal segment $n$
\ENSURE Nearest $s \in S$ w.r.t.\ $n$ or discard $n$
\STATE $(d_{\min}, s^*) \leftarrow (+\infty, \varnothing)$
\FOR{$s \in S$}
    \STATE $d \leftarrow \mathrm{distance}(n,s)$
    \IF{$n \cap s \neq \varnothing$}
        \STATE $(d_{\min}, s^*) \leftarrow (0, s)$
        \STATE \textbf{break}
    \ELSIF{$d < d_{\min}$}
        \STATE $(d_{\min}, s^*) \leftarrow (d, s)$
    \ENDIF
\ENDFOR
\IF{$d_{\min} > \delta$}
    \STATE discard $n$
\ELSE
    \STATE assign $s^*$ to $n$
\ENDIF
\end{algorithmic}
\end{algorithm}

To support non-verbal speech generation and understanding, it is necessary to integrate the textual content with non-verbal tags.
The related algorithm is presented in Algorithm~\ref{alg:insert_tag}, which implements the timestamp-based ordering (TBO) method.
It is worth noting that state-of-the-art ASR models—including Whisper-Large-V3~\cite{radford2023robust} and Paraformer-zh~\cite{gao2022paraformer}—offer word-level timestamps. In practice, however, we have observed that these timestamps exhibit varying degrees of misalignment, with Paraformer-zh showing particularly noticeable discrepancies. We try using forced alignment tools, such as Montreal Forced Aligner~\cite{mcauliffe2017montreal}, to improve alignment. However, due to insertion, substitution, and deletion errors in the transcriptions, no significant improvement is achieved; in some cases, the results are even worse.

\begin{algorithm}[htbp]
\centering
\caption{Insert Non-Verbal Tag into Transcription}
\label{alg:insert_tag}
\begin{algorithmic}[1]
\REQUIRE Non-verbal segment $n = [n_s, n_e]$; non-verbal label $l$; timestamped word list $W = \{(s_i, e_i, w_i)\}_{i=1}^N$
\ENSURE Updated word list with inserted non-verbal tags

\STATE $s\_tag \leftarrow ``[l]<B>"$
\STATE $e\_tag \leftarrow ``</B>"$
\STATE $i_s \leftarrow \max\{i \mid s_i \le n_s\}$
\STATE $i_e \leftarrow \min\{i \mid e_i \ge n_e\}$

\IF{$i_e - i_s > 1$}
    \STATE \# Spans multiple words
    \STATE \textbf{return} $W[0:i_s] + [s\_tag] + W[i_s:i_e] + [e\_tag] + W[i_e+1:N]$
\ELSE
    \STATE \# Single word
    \STATE \textbf{return} $W[0:i_e] + [$l$] + W[i_e+1:N]$
\ENDIF
\end{algorithmic}
\end{algorithm}

\section{NonVerbalSpeech-38K Dataset}

\begin{figure}[t!]
\centering
\includegraphics[width=0.85\columnwidth]{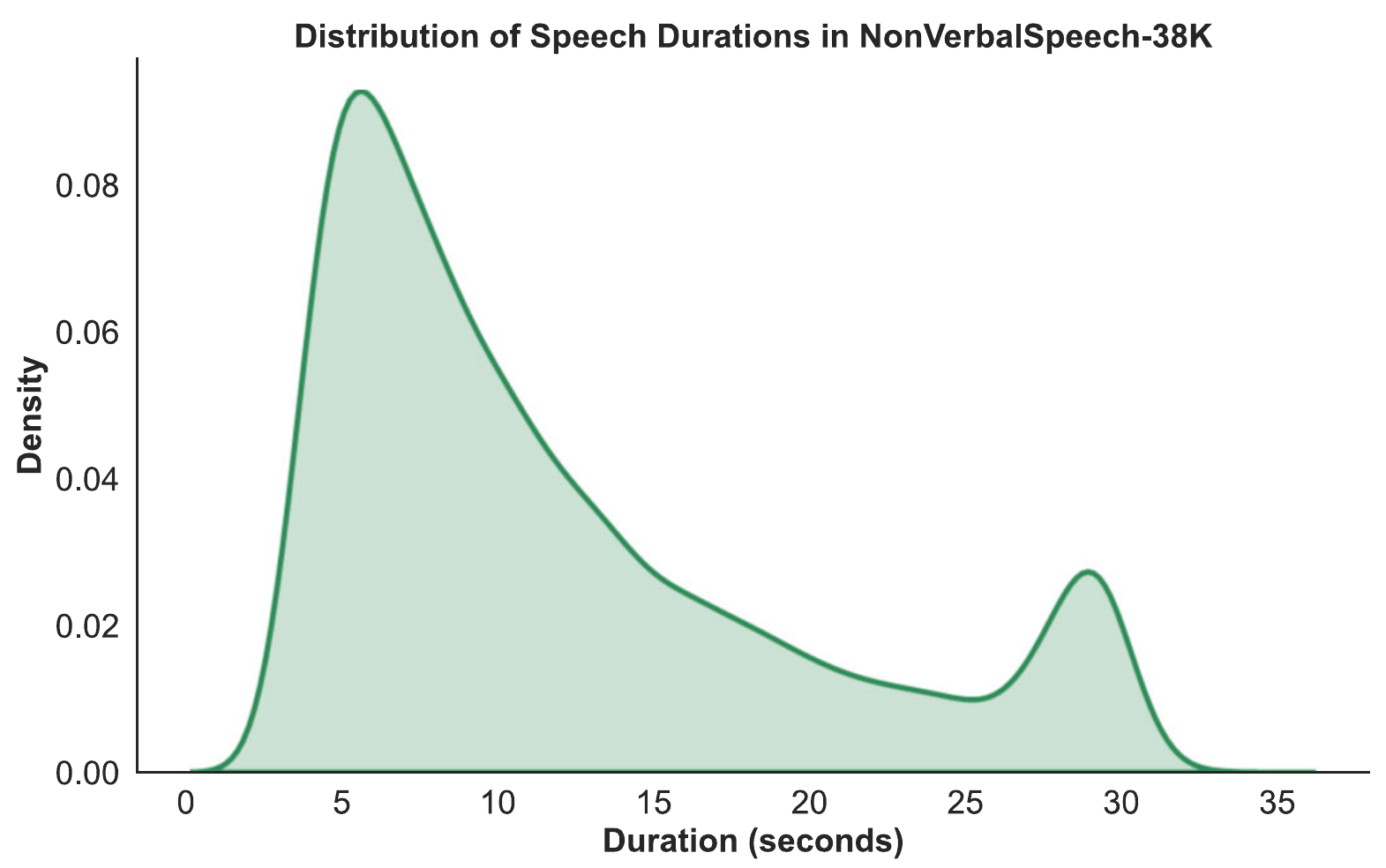}
\caption{Speech duration distribution of the proposed NonVerbalSpeech-38K dataset.}
\label{fig: dur-dist}
\end{figure}

\begin{table*}[t!]
\centering
\begin{threeparttable}
\caption{Source distribution in the original data and the NVS-38K dataset}
\footnotesize
\renewcommand{\arraystretch}{1.1}
\setlength{\tabcolsep}{2.5mm}
\begin{tabular}{lrrrrr}
\toprule
Source & Radio Dramas & Comedy Sketches & Cartoon & Variety Shows & Short Plays \\
\midrule
Crawled (hrs) & 17,400 & 7,200 & 3,600 & 1,400 & 1,200 \\
NVS-38K (\#) & 21,668 & 7,273 & 5,019 & 1,217 & 809 \\
\midrule
Source & Speeches & Documentaries & Movies & Audiobooks & Toy Unboxing \\
\midrule
Crawled (hrs) & 1,079 & 600 & 500 & 263 & 9 \\
NVS-38K (\#) & 158 & 105 & 1,090 & 1,375 & 4 \\
\bottomrule
\end{tabular}
\label{tab:source_hours_final_samples}
\begin{tablenotes}
\item ``Crawled (hrs)" shows the duration of the original crawled data; ``NVS-38K (\#)" shows the final sample counts in the proposed NonVerbalSpeech-38K dataset.
\end{tablenotes}
\end{threeparttable}
\end{table*}

Using the proposed NonVerbalSpeech-38K pipeline, we construct the NonVerbalSpeech-38K dataset from a large colillection of speech data crawled from diverse online sources (Table~\ref{tab:source_hours_final_samples}). 
This diversity ensures coverage of a wide range of natural speaking styles and, importantly, includes rich and varied non-verbal phonemes. 
Radio dramas, comedy sketches, and cartoons contribute the most data, while movies and audiobooks yield the highest processing conversion rates, suggesting richness in non-verbal content.
The resulting durations mostly fall within the range of 3 to 30 seconds (Figure~\ref{fig: dur-dist}), primarily because the Emilia~\cite{he2024emilia} pipeline segments speech clips into chunks of 3–30 seconds. This segmentation operation also explains the peak near 30 seconds. Some segments slightly exceeding 30 seconds arise during Step~3 of the NonVerbalSpeech-38K pipeline (Algorithm~\ref{alg:nearest_speech}), where non-verbal segments that do not fall within any speech region but are within a 1-second distance threshold are merged with the nearest speech region to maintain contextual continuity. 

\textbf{Limitations}: We observe a language imbalance in the current dataset, because the original source datasets are predominantly in Chinese, rather than due to limitations of our pipeline.
In fact, our frame-level detection model is trained exclusively on English data, yet it demonstrates strong generalization performance on the Chinese test set. Moreover, since the majority of samples in our dataset are in Chinese, this further substantiates the robustness and cross-lingual generalization ability of the proposed detection model. 
Thanks to our fully automatic and scalable pipeline, further expansion of the dataset is feasible, and future work will continue to increase both its size and language coverage, helping to mitigate the current language imbalance.

\textbf{License}: Our dataset has been publicly released with the following declaration: ``The NVS-38K dataset does not hold the copyrights of the audio files it contains; copyright ownership remains with the original creators. The dataset is made available solely for non-commercial research purposes under the Creative Commons Attribution–NonCommercial 4.0 International (CC BY-NC 4.0) license". We hope that releasing this dataset will facilitate further research on non-verbal speech.

\end{document}